# Collapse to Black Holes in Brans-Dicke Theory:
# II. Comparison With General Relativity

Mark A. Scheel

*Center for Radiophysics and Space Research
and Department of Physics,
Cornell University, Ithaca, New York 14853*

Stuart L. Shapiro and Saul A. Teukolsky

*Center for Radiophysics and Space Research
and Departments of Astronomy and Physics,
Cornell University, Ithaca, New York 14853*

November 1994

ABSTRACT: We discuss a number of long-standing theoretical questions about collapse to black holes in the Brans-Dicke theory of gravitation. Using a new numerical code, we show that Oppenheimer-Snyder collapse in this theory produces black holes that are identical to those of general relativity in final equilibrium, but are quite different from those of general relativity during dynamical evolution. We find that there are epochs during which the apparent horizon of such a black hole passes *outside* the event horizon, and that the surface area of the event horizon *decreases* with time. This behavior is possible because theorems which prove otherwise assume $R_{ab}l^a l^b \geq 0$ for all null vectors $l^a$. We show that dynamical spacetimes in Brans-Dicke theory can violate this inequality, even in vacuum, for any value of $\omega$.

## I. INTRODUCTION

Scalar-tensor theories of gravity differ from general relativity (GR) because they describe gravitation using not only a spacetime metric, but also a scalar field that couples to both matter and the spacetime geometry. These theories have recently regained popularity, in part because they arise naturally as the low-energy limit of many theories of quantum gravity, such as Kaluza-Klein theories[1] and supersymmetric string theories[2]. Scalar-tensor gravitation is also important for "extended" cosmological inflation models[3], in which the scalar field provides a natural termination of the inflationary era via bubble nucleation without the need for finely tuned cosmological parameters. In addition, inflation-induced oscillation of a massive gravitational scalar field has been considered as a candidate for the "missing mass" required to close the universe[4].

Scalar-tensor theories contain adjustable parameters that describe the coupling between the scalar field, matter, and the spacetime metric. For certain values of these parameters, post-Newtonian expansions of scalar-tensor theories agree with GR. For this reason, solar system observations and experiments cannot rule out scalar-tensor gravitation in favor of GR, but can only place limits on scalar-tensor coupling parameters. Currently, solar system experiments require[5]

$$|\zeta| < 10^{-3}, \qquad (1.1)$$

where $\zeta$ is the ratio of the coupling between matter and the scalar field to the coupling between matter and the metric. Such a small value of $\zeta$ seems to suggest that a gravitational scalar field does not exist. However, for a large class of scalar-tensor theories in which $\zeta$ depends on the scalar field, the expansion of the universe during the matter-dominated era naturally drives $\zeta$ towards zero[6]. In other words, many scalar-tensor theories that differ significantly from GR in the early universe become nearly indistinguishable from GR in the present epoch. Hence, the experimental evidence that supports GR need not be viewed as an argument against scalar-tensor gravitation.

Because scalar-tensor gravitation can agree with GR in the post-Newtonian limit, it is important to study strong-field examples in which the two theories may give different predictions. These examples may not only provide further experimental and observational tests that might distinguish between GR and scalar-tensor gravitation, but they may also illuminate the structure of both theories.

One such strong-field example is the generation of gravitational waves. Because scalar-tensor theories involve a scalar field, they allow spherical sources such as supernovae to emit monopole radiation and binary systems to emit dipole radiation[7]. In contrast, GR only allows modes with quadrupole and higher angular dependence. The extra polarization states admitted by scalar-tensor gravitation would not only result in different gravitational waveforms than in GR, but would also increase the total energy radiated by a given source over the amount predicted by GR. Indeed, it has been suggested[8] that gravitational wave detectors like LIGO may be capable of distinguishing between GR and scalar-tensor gravitation, or of setting more stringent limits on scalar-tensor parameters.

Another strong-field example in which a scalar field may be important is the formation of black holes and singularities during gravitational collapse. Scalar-tensor gravitation admits a larger number of solutions than GR because it involves more degrees of freedom. Even in spherically symmetric vacuum situations, a variety of both static and dynamical solutions are possible because Birkhoff's theorem does not hold. Some of these solutions have undesirable properties such as naked singularities. Rather than simply ignoring these solutions on physical grounds, we should attempt to determine whether or not they can be produced by physical, nonsingular initial data. If pathological solutions could be produced by gravitational collapse of an initially nonsingular matter distribution, this would reveal a fundamental flaw in the predictive power of scalar-tensor gravitation. In GR, it is believed (but has not been proven; see, *e.g.*, Ref. [9]) that all physical singularities are hidden inside event horizons (the "cosmic censorship conjecture" [10]), so that the spacetime geometry outside these horizons is nonsingular and predictable. Studying whether cosmic censorship holds in scalar-tensor theories may clarify the issue in GR.

We concentrate on Brans-Dicke (BD) theory[11], the simplest of the scalar-tensor theories. BD contains a massless scalar field $\phi$ and a single dimensionless constant $\omega$ that describes the strength of the coupling between $\phi$ and the matter. The post-Newtonian expansions of BD and GR agree in the limit $|\omega| \to \infty$. BD is consistent with solar system observations and experiments[5] for $|\omega| \gtrsim 500$. Although this relation permits both positive and negative values of $\omega$, it is usually assumed that $\omega > 0$ because positive values of $\omega$ result in a positive contribution of matter to the scalar field and a positive value of the scalar field's stress-energy. Accordingly, for the majority of this paper we restrict ourselves to nonnegative values of $\omega$. However, for completeness we consider scenarios with negative $\omega$ in Section V.B.

We have developed a new numerical code for Brans-Dicke theory that solves the coupled dynamical equations for the metric and scalar field for a spherically symmetric matter source consisting of collisionless particles. We use a traditional singularity-avoiding (SA) technique to solve the equations at early times in the simulation, and we employ an apparent horizon boundary condition (AHBC) method after the formation of a black hole. The latter method allows us to follow the evolution of the spacetime arbitrarily far into the future without encountering either coordinate pathologies or spacetime singularities. Our numerical method is described in detail in Ref.[12], henceforth referred to as Paper I. Here we use our code to address several long-standing questions concerning spherical collapse in Brans-Dicke theory.

Spherical BD collapse allows one to explore two strong-field phenomena in a simple setting: the generation of (monopole) gravitational radiation, and the formation of black holes and singularities. It is therefore not surprising that this problem has received much attention in the literature[13, 14, 15, 16]. However, because detailed studies of nonlinear time-dependent collapse were impeded by lack of numerical techniques, most treatments involve perturbation analyses or discussions of the final state of the collapsed object, as discussed in Section III. Only very recently[8] has there been any calculation of the time-dependent gravitational waveform produced by BD collapse. The dynamical behavior of the black hole produced by such an event has never been studied in detail.

We show numerically that for $\omega \geq 0$, Oppenheimer-Snyder collapse in BD ends in a black hole rather than a naked singularity, in agreement with the results of Shibata *et al.*[8]. We calculate accurate gravitational waveforms produced by the collapse, and evaluate the mass lost to gravitational radiation. At late times, we find that the exterior region of the resulting black hole is described by the Schwarzschild metric with a constant scalar field. This is expected, since a theorem of Hawking[16] states that *stationary* black holes in BD are identical to those in GR. However, we find that during the *dynamical* epoch in which it radiates mass, a BD black hole behaves quite differently from a black hole in general relativity. We find an epoch during which *the event horizon of a dynamical BD black hole passes inside the apparent horizon, and the surface area of the event horizon decreases in time*. This happens for all $\omega$, and is possible because the spacetime does not satisfy the null energy condition

$$R_{ab} l^a l^b \geq 0 \qquad \text{for all null vectors } l^a, \tag{1.2}$$

where $R_{ab}$ is the Ricci tensor.

For completeness, we also consider Oppenheimer-Snyder collapse for some negative values of $\omega$. As in the case of positive $\omega$, we find that the collapse ends in a black hole rather than a naked singularity, and that this black hole settles down to the Schwarzschild solution at late times. During dynamical epochs, a black hole with $\omega < 0$ violates the null energy condition for most of its evolution, and possesses an event horizon that decreases in area and lies inside the apparent horizon.

## II. PROPERTIES OF BRANS-DICKE THEORY

In Brans-Dicke theory, gravitation is described by a metric $g_{ab}$ and a scalar field $\phi$. The scalar field obeys a wave equation with a source term determined by the matter distribution. As in GR, test particles move on geodesics of the metric; they feel no additional force from the scalar field. However, unlike GR, the equations determining the metric are not Einstein's equations—they contain second derivatives of $\phi$.

### A. Basic Equations

The action for Brans-Dicke gravitation is[11]

$$I = \int \mathcal{L}_{\mathrm{BD}} (-g)^{1/2} d^4 x, \tag{2.1}$$

where the Lagrangian density is

$$\mathcal{L}_{\mathrm{BD}} = g^{ab} R_{ab} \phi + \frac{16\pi}{c^4} \mathcal{L} - \frac{\omega}{\phi} g^{ab} \partial_a \phi \partial_b \phi. \tag{2.2}$$

The coupling constant $\omega$ is dimensionless, and the scalar field $\phi$ has dimensions of $G^{-1}$, where $G$ is Newton's gravitational constant. The Lagrangian density $\mathcal{L}$ for matter and nongravitational fields depends on the metric $g^{ab}$ but not on $\phi$. The Ricci tensor $R_{ab}$ is obtained from the metric in the usual way.

We choose our units such that $c = \phi_\infty = 1$, where $\phi_\infty$ is the value of $\phi$ far from all sources. Because we must have[11]

$$\frac{1}{\phi_\infty} = \frac{3 + 2\omega}{4 + 2\omega} G \tag{2.3}$$

in order for Brans-Dicke theory to reproduce the measured value of the gravitational constant in the Newtonian limit, our choice of units implies

$$G = \frac{4 + 2\omega}{3 + 2\omega} \neq 1. \tag{2.4}$$

Variation of the action (2.1) with respect to $g^{ab}$ and $\phi$ yields the field equations

$$\Box \phi = \frac{8\pi T}{3 + 2\omega}, \tag{2.5}$$

$$G_{ab} \phi = 8\pi T_{ab} + \frac{\omega}{\phi} \left( \nabla_a \phi \nabla_b \phi - \tfrac{1}{2} g_{ab} \nabla_a \phi \nabla^a \phi \right) + \nabla_a \nabla_b \phi - g_{ab} \Box \phi. \tag{2.6}$$

Here $\nabla$ denotes covariant differentiation with repect to the metric, $\Box$ is the covariant Laplacian $\nabla^a \nabla_a$, and $G_{ab}$ is the usual Einstein tensor. The term involving $\omega$ on the right-hand side of Eq. (2.6) is the stress-energy of the scalar field. The last two terms in this equation ensure that both the contracted Bianchi identity $\nabla_b G^{ab} = 0$ and the energy-momentum conservation law $\nabla_b T^{ab} = 0$ remain satisfied even though Einstein's equations are not. These terms make Brans-Dicke theory different from general relativity with a Klein-Gordon scalar field for a source.

### B. The Einstein Representation

Eqs. (2.5) and (2.6) are written in a representation in which test particles have constant rest masses and move on geodesics, but the field equations differ from Einstein's equations. This is known as the canonical or Brans-Dicke representation. One can also work in the Einstein representation[17], a conformal frame in which the field equations (2.6) become Einstein's equations but test particles have rest masses proportional to $\phi^{-1/2}$.

One transforms to the Einstein representation by leaving the coordinates invariant and transforming the metric and stress-energy tensor according to

$$\bar{g}_{ab} = \phi G_0 g_{ab}, \tag{2.7a}$$

$$\bar{T}_{ab} = \phi^{-1} G_0^{-1} T_{ab}. \tag{2.7b}$$

This can be thought of as a position-dependent rescaling of the units of length, time, and reciprocal mass. The quantity $G_0$ is an arbitrary constant with dimensions of $G$. Note that $G_0$ merely represents a choice of mass units, and is not equal to the value of the gravitational constant measured, for example, by a Cavendish experiment. As in the Brans-Dicke representation, the locally measured gravitational constant depends on $\phi$, and therefore can vary in space and time.

The field equations resulting from Eqs. (2.7) are

$$\bar{\Box}(\ln \phi) = \frac{8\pi \bar{T}}{3 + 2\omega}, \tag{2.8}$$

$$\bar{G}_{ab} = 8\pi G_0 \bar{T}_{ab} + \frac{3 + 2\omega}{16\pi G_0 \phi^2} \left( \bar{\nabla}_a \phi \bar{\nabla}_b \phi - \tfrac{1}{2} \bar{g}_{ab} \bar{\nabla}_a \phi \bar{\nabla}^a \phi \right), \tag{2.9}$$

where $\bar{G}_{ab}$, $\bar{\nabla}$, and $\bar{\Box}$ are the Einstein tensor, covariant derivative, and wave operator obtained from the unphysical metric $\bar{g}_{ab}$. We have set $c = 1$.

The rest mass $\bar{m}$ of a test particle in the Einstein representation is

$$\bar{m} \equiv m \phi^{-1/2} G_0^{-1/2}, \tag{2.10}$$

where $m =$ constant is the particle's rest mass in the Brans-Dicke representation. The trajectory of such a particle is given by[17]

$$\frac{d}{d\bar{\tau}} \left( \bar{m} \bar{u}_a \right) - \tfrac{1}{2} \bar{m} \bar{g}_{bc,a} \bar{u}^b \bar{u}^c + \bar{m}_{,a} = 0, \tag{2.11}$$

where $\bar{\tau}$ is the proper time of the particle as measured in the Einstein representation, and commas represent partial derivatives. This expression differs from the geodesic equation because of the third term, which represents an external force due to the scalar field.

Eqs. (2.9) contain no second derivatives of $\phi$; they are simply Einstein's equations with a Klein-Gordon scalar field source. Because of this, the Einstein representation is useful in extending results from GR over to BD, particularly in vacuum. However, it is awkward for discussing physical situations because gravitational effects on matter are not produced solely by the metric: rest masses of particles depend on the scalar field. For example, Dicke[17] points out that the gravitational redshift of spectral lines is described in this representation as a combination of both a metric effect and a $\phi$-induced change of atomic and molecular energy levels.

The Brans-Dicke representation is more useful for physical interpretations because the rest masses of particles are constant and matter obeys the conservation law $\nabla_a T^{ab} = 0$. This makes it much easier easier to incorporate non-gravitational physics into the theory. Unless otherwise stated, we will work in the Brans-Dicke representation.

### C. Equivalence Principle

The weak equivalence principle states that the trajectory of an uncharged test particle with negligible self-gravity is independent of all properties of the particle itself. BD satisfies the weak equivalence principle because it is a metric theory: worldlines of freely falling test particles are determined by the metric alone. However, BD does not satisfy the strong equivalence principle, which requires that the weak equivalence principle be satisfied even for bodies with large gravitational self-energy.

As an extreme case of equivalence principle violation, consider a small matter particle and a small Schwarzschild black hole moving through a background spacetime. In this case, the analysis is clarified by using the Einstein representation. As discussed earlier, a small matter particle does not move on a geodesic in this representation because its rest mass varies with the scalar field. A Schwarzschild black hole, on the other hand, is purely a metric phenomenon and contains no matter. Because the metric in this representation obeys

Einstein's equations, the black hole moves on a geodesic, as in general relativity[16]. Thus, in Brans-Dicke theory, a small Schwarzschild black hole and a small matter particle move on different trajectories.

Similarly, the trajectory of a massive particle in Brans-Dicke theory depends on the ratio of its gravitational binding energy to its total energy (the Nordtvedt effect[18]). Even in the Newtonian approximation of BD, massive bodies possess an additional non-Newtonian $1/r^2$ acceleration towards an external mass[18]. In other words, a massive object in Brans-Dicke theory has a passive gravitational mass greater than its inertial mass.

### D. Masses

The general asymptotically flat, static solution to the Brans-Dicke equations in spherical symmetry can be written (see Sections II.D and II.E of Paper I)

$$g_{00} = -1 + \frac{2M_{\rm T} + 2M_{\rm S}}{r}, \tag{2.12a}$$

$$g_{0i} = 0, \tag{2.12b}$$

$$g_{ij} = 1 + \delta_{ij}\left(\frac{2M_{\rm T} - 2M_{\rm S}}{r}\right), \tag{2.12c}$$

$$\phi = 1 + \frac{2M_{\rm S}}{r}. \tag{2.12d}$$

The quantities $M_{\rm S}$ and $M_{\rm T}$ are constants, and are known as the scalar and tensor masses[19]. From Eq. (2.12a), the Keplerian mass (active gravitational mass) measured by a test particle is

$$M \equiv M_{\rm T} + M_{\rm S}. \tag{2.13}$$

The scalar mass $M_{\rm S}$ is so named because it describes the $1/r$ dependence of the scalar field. The tensor mass $M_{\rm T}$ is the Keplerian mass (active gravitational mass) measured by a test Schwarzschild black hole in the asymptotic region of the spacetime. This is easily seen by transforming Eqs. (2.12) into the Einstein representation, in which Schwarzschild black holes move on geodesics: using the transformation (2.7) and retaining only first-order terms, we obtain

$$\bar{g}_{00} = -1 + \frac{2M_{\rm T}}{r}, \tag{2.14a}$$

$$\bar{g}_{0i} = 0, \tag{2.14b}$$

$$\bar{g}_{ij} = 1 + \delta_{ij}\left(\frac{2M_{\rm T}}{r}\right), \tag{2.14c}$$

$$\phi = 1 + \frac{2M_{\rm S}}{r}. \tag{2.14d}$$

Eq. (2.14a) shows that Keplerian orbits in the Einstein frame indeed measure a mass $M_{\rm T}$.

The appearance of two free parameters, $M_{\rm T}$ and $M_{\rm S}$, in the asymptotic solution (2.12) is related to the violation of the equivalence principle. In general, the ratio of $M_{\rm S}$ to $M_{\rm T}$ depends on the gravitational binding energy of the source. For example, the asymptotic metric of a Schwarzschild black hole of Keplerian mass $M$ is given by Eqs. (2.12) with

$$M_{\rm T} = M, \tag{2.15a}$$

$$M_{\rm S} = 0, \tag{2.15b}$$

but the external solution for a stationary, weakly gravitating spherical body of rest mass $m$ is given by Eqs. (2.12) with[11]

$$M_{\rm T} = m, \tag{2.16a}$$

$$M_{\rm S} = m\frac{1}{3+2\omega}, \tag{2.16b}$$

$$M = m\frac{4+2\omega}{3+2\omega}. \tag{2.16c}$$

Note that the solution (2.16) assumes that the gravitational field is weak even inside the source, so it is valid for an object like the sun but not for a neutron star. In addition, although Eq. (2.16c) indicates that the Keplerian mass of a weak source is different from its rest mass, it is important to note that Kepler's law for the metric (2.12) reads (restoring units using $\phi_\infty$)

$$\Omega^2 r^3 = (M_{\rm T} + M_{\rm S})\phi_\infty^{-1} = M\phi_\infty^{-1}, \tag{2.17}$$

where $\Omega$ is the orbital frequency. Substituting Eqs. (2.16c) and Eq. (2.3), we see that for a weak source we recover the Newtonian result $\Omega^2 r^3 = mG$. For strong sources such as neutron stars, the quantity measured by Kepler's law is still $M\phi_\infty^{-1}$, but this quantity is not simply related to the rest mass of the source because there is significant internal energy and gravitational binding energy.

The quantity most similar to the usual concept of mass is not the Kepler mass $M = M_{\rm T} + M_{\rm S}$, but the tensor mass $M_{\rm T}$. Indeed, with the natural choice of units $\phi_\infty = 1$, it is $M_{\rm T}$ and not $M$ that corresponds to the rest mass of a weakly gravitating body (Eq. (2.16)). Like the ADM mass in GR, the tensor mass is positive definite, decreases monotonically by emission of gravitational radiation, and is well-defined even for dynamical spacetimes[19]. The scalar mass $M_{\rm S}$ and the Kepler mass $M$ have none of these properties.

Conservation laws for tensor and scalar masses are obtained from the relation[19]

$$\phi^{n-1}\left(U^{\mu\nu} + T^{\mu\nu}\right) = \frac{1}{16\pi}\left[\phi^n(-g)\left(g^{\mu\nu}g^{\alpha\beta} - g^{\mu\alpha}g^{\nu\beta}\right)\right]_{,\alpha\beta}, \tag{2.18}$$

which implies

$$\left[\phi^{n-1}\left(U^{\mu\nu} + T^{\mu\nu}\right)\right]_{,\nu} = 0 \tag{2.19}$$

because of antisymmetry of the indices $\alpha$ and $\nu$ on the right-hand side of Eq. (2.18). Here $T^{\mu\nu}$ is the stress-energy tensor, $n$ is an arbitrary integer, and $U^{\mu\nu}$ is a pseudotensor that involves first derivatives of the metric and first and second derivatives of the scalar field. The pseudotensor $U^{\mu\nu}$ is different for each value of $n$. Explicit expressions for $U^{\mu\nu}$ are given in Ref. [19]. One recovers the GR limit by setting the scalar field equal to a constant. In this case, Eq. (2.18) becomes independent of $n$, and $U^{\mu\nu}$ reduces to the Landau-Lifshitz pseudotensor[20] multiplied by the factor $\phi/16\pi$.

From Eq. (2.18) one can obtain a conserved "mass" for any value of $n$. Let

$$\mathcal{M}(r;n) \equiv \frac{1}{16\pi}\int\left[(-g)\phi^n\left(g^{00}g^{ij} - g^{0i}g^{0j}\right)\right]_{,j}d^2\Sigma_i, \tag{2.20}$$

where $\Sigma_i$ is the two-dimensional area element on a sphere of radius $r$ in the asymptotic rest frame of the source, and $i,j$ refer to components in Cartesian coordinates. Then define

$$\mathcal{M}(n) \equiv \lim_{r\to\infty}\mathcal{M}(r;n). \tag{2.21}$$

For stationary spacetimes, one can show that

$$\mathcal{M}(n) = M_{\rm T} - \frac{1}{2}(2-n)M_{\rm S} \tag{2.22}$$

by inserting Eqs. (2.12) into Eq. (2.20). Following Lee[19], we therefore define the tensor mass of an arbitrary (possibly dynamical) asymptotically flat spacetime by

$$\mathcal{M}_{\text{T}} \equiv \mathcal{M}(2), \tag{2.23}$$

which reduces to $M_{\text{T}}$ in the stationary case.

Similarly, we can identify the scalar mass with the quantity

$$\mathcal{M}_{\text{s}}(n) \equiv \frac{2}{2-n} \left( \mathcal{M}(2) - \mathcal{M}(n) \right), \qquad n \neq 2, \tag{2.24}$$

because $\mathcal{M}_{\text{s}}(n) = M_{\text{s}}$ for a stationary spacetime. However, this quantity is not unique or even well-defined for a time-dependent spacetime: for $n \neq 2$, $\mathcal{M}(r; n)$ will in general depend on both $r$ and $n$, and $\lim_{r \to \infty} \mathcal{M}(r; n)$ will not exist. This is easily seen for the example of a dynamical spherically symmetric spacetime, in which

$$\mathcal{M}(r; n) = M_{\text{T}} + \frac{n-2}{4} \left( f(t-r) + r f'(t-r) \right) \tag{2.25}$$

for large $r$. Here $f$ is an arbitrary function of $(t - r)$ and $f'$ is its derivative. Eq. (2.25) can be derived using the asymptotic metric (I.2.53)–(I.2.56), where the prefix I refers to an equation in Paper I. For $n \neq 2$, this function depends on both $r$ and $n$, and does not approach a limit as $r \to \infty$. In contrast, the tensor mass (2.23) in this case is unique since $\mathcal{M}(r; 2) = M_{\text{T}}$ = constant, independent of $r$.

Despite the impossibility of defining a dynamical scalar mass, the net change in scalar mass $\Delta M_{\text{s}}$ from a system that is initially stationary, emits gravitational radiation, and settles down into a final equilibrium state, *is* a unique, well-defined quantity[19]. For sufficiently large $r$ and any $n \neq 2$,

$$\Delta M_{\text{s}} = \mathcal{M}_{\text{s}}(r; n)_{final} - \mathcal{M}_{\text{s}}(r; n)_{initial}. \tag{2.26}$$

We will therefore use the "instantaneous scalar mass function"

$$\mathcal{M}_{\text{s}}(r) \equiv \mathcal{M}_{\text{s}}(r; 0) \tag{2.27}$$

to track the change in scalar mass during dynamical epochs. However, keep in mind that this quantity has no unique physical meaning except in time-independent situations; only changes in $\mathcal{M}_{\text{s}}(r)$ from one stationary state to another are physically measurable.

Lee[19] has shown that if the scalar mass of a spherical system changes by an amount $\Delta M_{\text{s}}$ in a time $\tau$ (measured at infinity), the system must also lose an amount of tensor mass given by

$$\Delta M_{\text{T}} \geq (3 + 2\omega) \frac{(\Delta M_{\text{s}})^2}{\tau}. \tag{2.28}$$

### E. Spherically Symmetric Vacuum Solution

Brans[21] has constructed the exact static vacuum solution to Eqs. (2.6) and (2.5) in spherical symmetry. This solution can take one of four possible forms, depending on the values of arbitrary constants appearing in the solution. The Brans type I solution is the only form that is permitted for all values of $\omega$; the other three forms are only allowed for $\omega \leq -3/2$. We write the Brans type I solution using an isotropic spatial coordinate:

$$ds^2 = -\xi^{2(Q-\chi)} dt^2 + \left(1 + \frac{r_0}{r}\right)^4 \xi^{2(1-Q)} \left(dr^2 + r^2 \, d\Omega^2\right), \tag{2.29}$$

$$\phi = \xi^{\chi}, \tag{2.30}$$

where

$$\xi \equiv \frac{r - r_0}{r + r_0}, \tag{2.31}$$

and $r_0$, $Q$, and $\chi$ are arbitrary constants subject to the constraint

$$Q^2 + \chi^2 \left(1 + \frac{\omega}{2}\right) - Q\chi - 1 = 0. \tag{2.32}$$

The quantity $r_0$ is a mass parameter: it is related to the scalar and tensor masses by

$$M_{\rm S} = -r_0 \chi, \tag{2.33}$$
$$M_{\rm T} = r_0(2Q - \chi). \tag{2.34}$$

Our parameters $\chi$ and $Q$ correspond to $C/\lambda$ and $(1+C)/\lambda$ in Brans' notation. We have eliminated a constant conformal factor and the arbitrary constant associated with the choice of time coordinate by requiring the metric (2.29) to be asymptotically Minkowskian. For $\chi = 0$ and $Q = 1$, this solution reduces to the Schwarzschild solution with $\phi =$ constant for any value of $\omega$.

Campanelli and Lousto[22] have examined this metric using a different parameterization, and have found that it is asymptotically flat for all $Q$ and $\chi$, and that the surface $r = r_0$ acts as an event horizon when $2Q - \chi > 1$. In addition, the scalar invariant $I = R^{abcd}R_{abcd}$ is finite at $r = r_0$ whenever $Q = 1, \chi = 0$ (the Schwarzschild solution), or whenever $Q \geq 2$. In the latter case, $\omega \to -\infty$ as $\chi \to 0$, so this metric agrees with the Schwarzschild solution in the post-Newtonian limit. This solution is not merely a relabeling of the Schwarzschild metric because some components of the Ricci tensor are nonzero[22], even in the limit $\chi \to 0$. In addition, although this solution possesses a nonsingular event horizon ($I$ is finite), it is peculiar because the area of this horizon is infinite ($g_{\theta\theta}$ diverges).

### F. Limiting cases

It is often stated that Brans-Dicke theory reduces to general relativity in the limit $|\omega| \to \infty$. This is not entirely correct. It is certainly true that any solution of Einstein's equations is also a solution of the Brans-Dicke field equations (2.6) with $\phi$ strictly constant, and that $\phi =$ constant is a solution of the wave equation (2.5) for $|\omega| = \infty$. However, this by no means implies that *all* Brans-Dicke solutions satisfy Einstein's equations in the limit $|\omega| \to \infty$ or in the limit $\phi \to$ constant.

For example, although the Brans type I metric (2.29) reduces to the Schwarzschild solution for $\chi = 0$ and $Q = 1$, Matsuda[14] points out that if one assumes a finite $\omega$ and takes the limit $\chi \to 0$, $Q \to 1$, the areal radius

$$r_s = r \left(1 + \frac{r_0}{r}\right)^2 \xi^{(1-Q)} \tag{2.35}$$

tends to zero for $r = r_0$, but approaches $2r_0$ for $r = r_0 + \epsilon$ for any $\epsilon > 0$. This result is the truncated Schwarzschild metric[23], which contains a singular event horizon.

In addition, taking the limit $\chi \to 0$ with $Q \neq 1$ results in $|\omega| \to \infty$ and $\phi \to$ constant, but produces a metric that is different from the Schwarzschild solution, even though the field equations Eq. (2.6) with a constant scalar field reduce to Einstein's equations. This is because in these solutions derivatives of $\phi$ vanish like $\omega^{-1/2}$ rather than $\omega^{-1}$ as $|\omega| \to \infty$, and therefore the second term on the right-hand side of Eq. (2.6) approaches a finite value.

Although it is not mathematically rigorous to say that Brans-Dicke theory reduces to general relativity in the limit $|\omega| \to \infty$, such a statement may be correct for physical situations. The reason is that $|\omega| \to \infty$ solutions other than those of GR tend to have unphysical properties. For example, many solutions of Eq. (2.29) in the limit $\chi \to 0$ possess naked singularities. However, rather than simply discarding these unphysical solutions, it is important to determine whether they can be produced by nonsingular initial data.

There is an additional reason why solutions with $|\omega| \to \infty$ different from those of GR may not be physically relevant. In the presence of matter, regularity conditions require a static spherically symmetric spacetime to obey

$$\frac{\phi_{,r_s}}{\phi} = \frac{8\pi}{3\phi_0} \frac{T_0 r_s}{3+2\omega} \tag{2.36}$$

near the origin[24]. Here $r_s$ is the areal radius, $\phi_0$ is the central value of $\phi$, and $T_0$ is the central value of the trace of $T_{ab}$. This boundary condition requires $\phi_{,r_s} \sim \omega^{-1}$ as $|\omega| \to \infty$. Since the above non-Schwarzschild $|\omega| \to \infty$ solutions require $\phi_{,r_s} \sim \omega^{-1/2}$ as $|\omega| \to \infty$, these solutions cannot serve as an exterior metric for a nonsingular spherical material body. This suggests that the collapse of such a body in the limit $|\omega| \to \infty$ should result in a Schwarzschild black hole.

## III. GRAVITATIONAL COLLAPSE

The subject of gravitational collapse in Brans-Dicke theory has been discussed extensively in the literature. Because there exist Brans-Dicke solutions different from those of GR, it had been conjectured[13] that a collapsing body reaches a different final state in the two theories. However, Penrose[25] suggested that the opposite is true: the end result of collapse in BD, as in GR, is either a Schwarzschild or Kerr black hole.

### A. Large $\omega$ theory[15]

Evidence for Penrose's conjecture was supplied by Thorne and Dykla[15], who investigated the question using an approximate version of BD for large values of $\omega$. In this formalism, one expands the metric, scalar field, and stress-energy tensor in powers of $1/\omega$, and drops higher-order terms in the field equations. The zeroth order terms yield Einstein's equations and a constant scalar field, so that the background metric is given by general relativity. The first order terms govern the evolution of Brans-Dicke perturbations about this general relativistic background.

Because the background metric is determined by general relativity, the only vacuum black hole solution with a nonsingular event horizon is the Kerr metric, with the Schwarzschild metric as a special case. For a Kerr background with $|a| < M$, theorems of Carter[26] and of Fackerell and Ipser[27] require that the scalar field must be constant and the metric perturbation must vanish everywhere if all physical quantities are to remain regular on the Kerr horizon and at infinity. Thus, the only black hole vacuum solution to the large $\omega$ theory without a naked singularity is the Kerr metric with a constant scalar field.

Furthermore, one can apply results of Price[28] to show that in the large $\omega$ theory, spherically symmetric collapse results in a Schwarzschild black hole. Any scalar field perturbation present in the initial data radiates away until $\phi = $ constant.

Thus, black holes without naked singularities in the large $\omega$ theory are identical to those in general relativity, and these black holes, at least in the nonrotating case, are produced by gravitational collapse of matter. This suggests that the same may be true for Brans-Dicke theory.

However, there are a few points that are not addressed by this analysis. Although the large $\omega$ theory is intended as an approximation to Brans-Dicke theory for experimentally acceptable values of $\omega$, the background metric must be a solution of Einstein's equations, and not one of the other $|\omega| \to \infty$ solutions, discussed in Section II.C, that are permitted by the Brans-Dicke equations. By expanding in $1/\omega$, one automatically excludes those solutions in which scalar field derivatives behave like $\omega^{-1/2}$. It is therefore not surprising that gravitational collapse in large $\omega$ theory produces black holes identical to those in general relativity, since the background metric excludes many other possibilities from the start. Furthermore, by assuming regularity at the event horizon of the final black hole, this analysis sidesteps the question of whether a naked singularity can result from collapse of a nonsingular object.

### B. Hawking's theorem

By working in the Einstein representation, Hawking[16] extended some of his theorems for general relativistic black holes to Brans-Dicke theory. In particular, if the Einstein-representation Ricci tensor satisfies the null energy condition

$$\bar{R}_{ab} l^a l^b \geq 0 \qquad \text{for all null } l^a, \tag{3.1}$$

then a stationary black hole must be either static or axisymmetric, and must have spherical topology. From this, Hawking proved that in such a solution the scalar field must be strictly constant. This implies that the black hole is a solution of Einstein's equations, and that the scalar mass $M_s$ is zero. Therefore, as long as Eq. (3.1) is satisfied, any object that collapses to a black hole in Brans-Dicke theory must radiate away all of its scalar mass before it settles into final equilibrium, and this equilibrium state will be either a Kerr or Schwarzschild spacetime. One can show from Eq. (2.9) that Eq. (3.1) will be satisfied as long as $\omega > -3/2$ and as long as the matter satisfies $\bar{T}_{ab} l^a l^b \geq 0$.

Note that Hawking's theorem assumes the existence of a black hole with a regular event horizon; it does not address the question of whether Brans-Dicke collapse proceeds instead to a naked singularity. Although this possibility is unphysical, it is interesting to consider whether it is automatically excluded by the theory.

### C. Numerical Simulations

Before the modern development of numerical relativity, Matsuda and Nariai[29] numerically evolved adiabatic spherical BD collapse of an ideal gas. They found that for $\omega = 10$, the early features of the collapse are not too different from the GR result. However, since they used for their evolution a generalization of the Misner-Sharp[30] equations, which become singular at an apparent horizon, they could not determine the end result of the collapse. In addition, because they used a Lagrangian hydrodynamics scheme covering only the matter, they were not able to treat propagation of monopole gravitational radiation into the vacuum region surrounding the collapsing object.

Recently, Shibata *et al.*[8] have simulated the spherical collapse of a pressureless fluid in Brans-Dicke theory for $\omega \geq 5$, and have calculated gravitational waveforms and spectra that result from such a collapse. They have found that an apparent horizon forms, and that the scalar field approaches a constant value afterwards, in agreement with Hawking's theorem.

### IV. NUMERICAL METHOD

We use a spherically symmetric mean-field particle simulation scheme to solve the Brans-Dicke equations for collisionless matter. We adopt the spatially isotropic metric

$$ds^2 = -(\alpha^2 - A^2 \beta^2) \, dt^2 + 2A^2 \beta \, dr \, dt + A^2 (dr^2 + r^2 \, d\Omega^2), \tag{4.1}$$

where $A$, $\alpha$, and $\beta$ are functions of $t$ and $r$, and we use the maximal time slicing condition

$$K^a_a = 0, \tag{4.2}$$

where $K^a_b$ is the extrinsic curvature tensor. All calculations are done in the physical Brans-Dicke representation of the theory, rather than in the Einstein conformal frame.

We calculate matter source terms by binning a finite number of particles into zones on a numerical grid. The source terms are then used to solve the field equations for the metric and $\phi$ by finite differencing. At each time step, particles are moved according to the geodesic equations. The maximal slicing condition prevents our code from encountering the spacetime singularity that forms at the origin after collapse to a black hole.

After an apparent horizon (AH) forms, we switch to an apparent horizon boundary condition (AHBC) scheme that is capable of integrating arbitrarily far into the future. By locking the AH to a fixed radial coordinate, we solve for the numerical variables in the region outside the AH, and discard the interior region. This is possible because the interior cannot causally influence the exterior. In the case of uniform or nearly uniform collapse, we can wait until all particles have fallen into the black hole before using the AHBC method; in this way, we do not need to include particles in our AHBC code.

Details of our numerical code are presented in Paper I.

## V. NUMERICAL RESULTS

### A. Oppenheimer-Snyder Collapse for $\omega \geq 0$

In this section we examine Oppenheimer-Snyder collapse in Brans-Dicke theory for nonnegative values of the coupling constant $\omega$. In these examples, an initially stationary uniform particle distribution of tensor mass $M_T(0)$ and areal radius $R_s = 10 M_T(0)$ is allowed to collapse to a final state. A black hole is formed in each case. Initially the spacetime is evolved using our SA method described in Paper I. After the formation of an apparent horizon, which occurs at about $t \sim 45 M_T(0)$, we switch to our AHBC method. By this time, all matter has already fallen into the black hole.

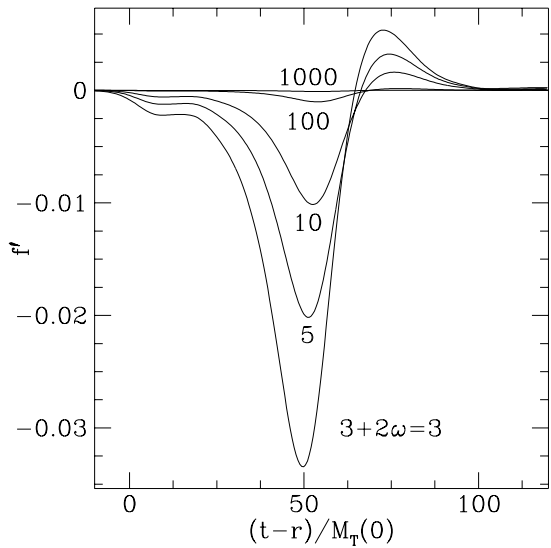

**Figure 1**  Gravitational waveform $f'(t - r)$ for Oppenheimer-Snyder collapse as seen by an observer far from the source, shown for five values of $\omega$. For each case, $M_T(0)/R_s = 0.1$.

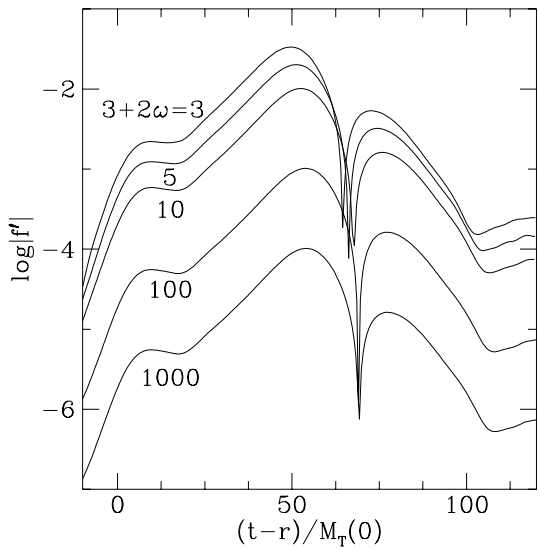

**Figure 2**  Same as Figure 1 except we plot the quantity $\log|f'(t-r)|$ in order to better show the behavior of the waveforms as $\omega \to \infty$. For large $\omega$, the wave amplitude is proportional to $1/(3+2\omega)$.

1. Gravitational radiation

The most obvious qualitative difference between BD and general relativity is that because BD contains a scalar field, it allows gravitational radiation even in spherical symmetry. The general spherically symmetric solution of the BD equations in the asymptotic region far from the source can be written (c.f. Eqs. I.2.53–I.2.56)

$$g_{00} = -1 + \frac{2M_{\rm T}}{r} + \frac{f(t-r)}{r}, \tag{5.1a}$$

$$g_{0i} = 0, \tag{5.1b}$$

$$g_{ij} = 1 + \delta_{ij}\left(\frac{2M_{\rm T}}{r} - \frac{f(t-r)}{r}\right), \tag{5.1c}$$

$$\phi = 1 + \frac{f(t-r)}{r}, \tag{5.1d}$$

where $f$ is an arbitrary function of $(t-r)$. For static situations, $f$ is the scalar mass $M_{\rm s}$ of the system. In Figures 1 and 2 we plot the gravitational wave amplitude $f'(t-r)$ for Oppenheimer-Snyder collapse, calculated by reading off the value of $r\partial\phi/\partial t$ far from the black hole at $r = 80M_{\rm T}(0)$. Here a prime denotes a deriviative with respect to the argument. The quantity $f'(t-r)$ depends only on $\omega$ and on the ratio $M_{\rm T}(0)/R_s$. For large $\omega$, we find that the gravitational wave amplitude is proportional to $1/(3+2\omega)$, in agreement with the results of Shibata[8] et al.. In the limit $\omega \to \infty$ no radiation is emitted, as in general relativity.

As the collapsing object emits monopole gravitational radiation, the tensor mass, scalar mass, and active gravitational mass decrease in time. Figure 3 shows the scalar mass as a function of time, normalized to the initial tensor mass. The resulting black hole radiates away all its scalar mass during the course of its evolution, in agreement with the results of Hawking[16] and of Shibata[8] et al.. This is equivalent to the statement that the scalar field approaches a constant value as $t \to \infty$. For the five values of $\omega$ shown, the net loss in scalar mass is $\Delta M_S/M_T(0) = 0.31, 0.19, 0.094, 0.0094$, and $0.00094$, in order of increasing $\omega$. We see that for large $\omega$, the total radiated scalar mass is proportional to $1/(3 + 2\omega)$.

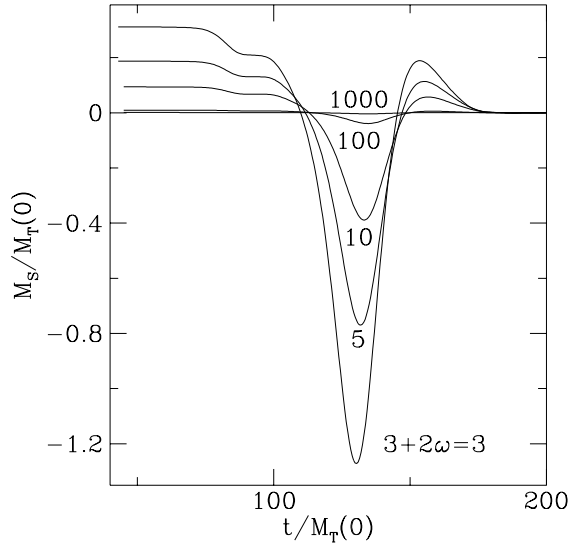

**Figure 3** Scalar mass function $\mathcal{M}_S(r)$ versus time for Oppenheimer-Snyder collapse, calculated at $r = 80 M_T(0)$ from Eq. (I.2.64.b) and shown for five values of $\omega$. The quantity $\mathcal{M}_S(r)$ is well-defined and unique only for stationary systems, so only the initial and final values of $\mathcal{M}_S(r)$ are physically meaningful, and are equal to the scalar mass $M_S$. The scalar mass is zero in the final state.

Figure 4 shows the tensor mass of the system as a function of time, calculated from Eq. (I.2.64a). Unlike the scalar mass $M_S$ and the active gravitational mass $M$, the tensor mass $M_T$ is a meaningful quantity even during dynamical epochs, and can be interpreted as the energy contained in the system. The tensor mass should decrease monotonically as the system loses energy to gravitational radiation. The reason this is not true in our simulations is because of higher order terms that are not included in Eq. (I.2.64a). These terms introduce $O(1/r)$ corrections to the tensor mass that would not be present if we evaluated Eq. (I.2.64a) at a much larger radius. These terms are not present in the final state of collapse (See Eq. I.5.7), and are more apparent in Figure 4 than in Figure 3 because the change in tensor mass is much smaller than the change in scalar mass.

By extrapolating the measured value of $\mathcal{M}_T(r)$ at several different radii to $r = \infty$ during the initial and final stationary states, we can determine the true change in tensor mass for the collapse to reasonable accuracy. This process, which is described in more detail in Paper I, yields $\Delta M_T/M_T(0) = 0.012, 0.006, 0.003, 0.0003$, and $0.00004$ for the five values of $\omega$ shown in Figure 4. For large $\omega$, our results indicate that the total radiated tensor mass is proportional to $1/(3 + 2\omega)$. For $\tau \sim 30 M_T(0)$, which from the figures is a reasonable lower limit on the collapse timescale, the quantity $(3 + 2\omega)(\Delta M_S)^2/(\tau \Delta M_T)$ is equal to 0.8 for $3 + 2\omega = 3$, and is very close to 1 for the other values of $\omega$. We see therefore that Lee's inequality (2.28) is satisfied.

The active gravitational mass $M$, normalized to the initial tensor mass $M_T(0)$, is shown in Figure 5. Although the tensor mass loss, *i.e.*, the total energy of the emitted gravitational radiation, is small, the change in $M$ can be much larger because of the large change in scalar mass. Although $M$ should not be interpreted as the energy of the system, it is the quantity measured by test particles in Keplerian orbits in the asymptotic region.

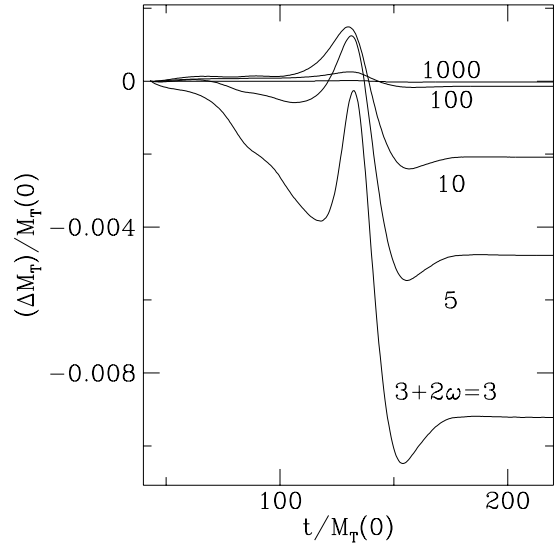

**Figure 4** Change in tensor mass function $\mathcal{M}_T$ versus time for Oppenheimer-Snyder collapse, calculated at $r = 80 M_T(0)$ using Eq. (I.2.64a) and shown for five values of $\omega$.

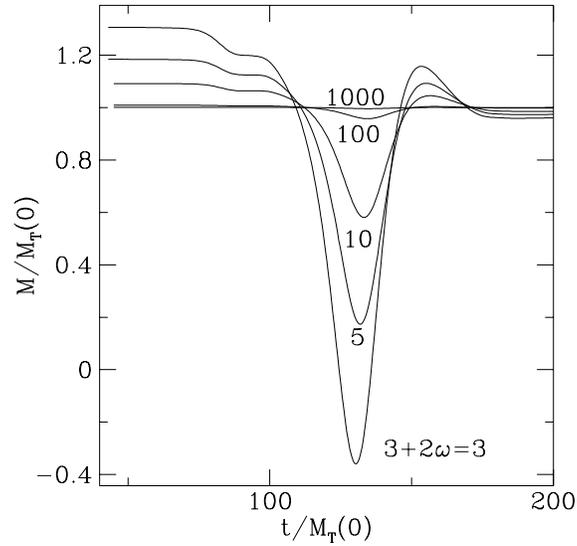

**Figure 5** Active gravitational mass $M(r) = \mathcal{M}_S(r) + \mathcal{M}_T$ versus time for Oppenheimer-Snyder collapse, measured at $r = 80 M_T(0)$ and shown for five values of $\omega$. The quantity $M$ is well-defined and unique only for stationary systems, so only the initial and final values of $M$ are physically meaningful.

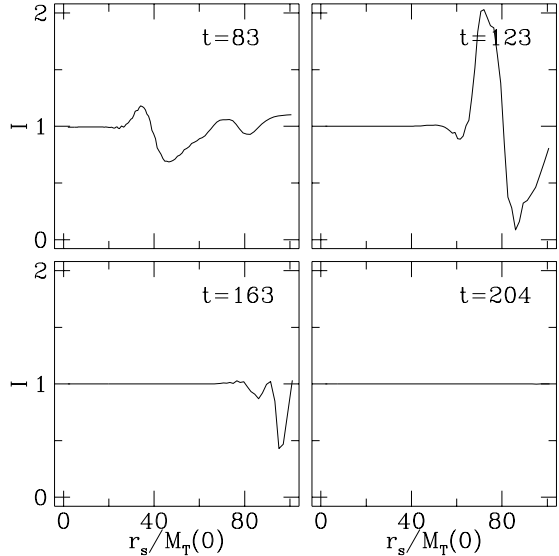

**Figure 6** The Riemann invariant $I$ in units of $48M_T^2 r_s^{-6}$ versus areal radius $r_s$, for Oppenheimer-Snyder collapse with $\omega = 0$ and initial areal radius $R_s = 10M_T(0)$. One can see the outgoing pulse of radiation before the black hole settles down into its stationary state. This state is the Schwarzschild solution, for which $M = M_T$ and $I/(48M_T^2 r_s^{-6}) = 1$.

### 2. Properties of Brans-Dicke Black Holes

According to Hawking's theorem[16], a stationary BD black hole with a nonsingular event horizon must satisfy Einstein's equations, so that a black hole produced by spherical BD Oppenheimer-Snyder collapse must obey the Schwarzschild solution once it settles down into a final stationary state. We show this explicitly by computing the scalar invariant

$$I = R^{abcd}R_{abcd}, \tag{5.2}$$

which is equal to $48M^2 r_s^{-6}$ for the Schwarzschild metric. We plot $I/(48M_T^2 r_s^{-6})$ versus areal radius in Figure 6 at several different times, for the collapse with $\omega = 0$. As the outgoing pulse of gravitational radiation propagates towards infinity, the black hole approaches the Schwarzschild solution. At $t = 500M_T(0)$, where we choose to terminate our numerical integration, the quantity $I/(48M_T^2 r_s^{-6})$ is equal to unity within two parts in $10^4$.

Figures 7 and 8 show the apparent horizons and event horizons of black holes resulting from BD Oppenheimer-Snyder collapse for different values of $\omega$. The event horizons are calculated by integrating outgoing null geodesics backwards in time, starting just outside the apparent horizon at $t = 200M_T(0)$. All such geodesics that are sufficiently close to the event horizon at $t = 200M_T(0)$ converge to the event horizon as they are followed backwards in time (they become null generators of the horizon). At final equilibrium, both the event horizon and apparent horizon lie at $r_s = 2M_T$, where the final tensor mass $M_T$ is less than its initial value $M_T(0)$ because energy has been carried to infinity by gravitational radiation. As $\omega \to \infty$, one recovers the result of GR: because the metric is Schwarzschild, the apparent and event horizons remain at $r_s = 2M_T = 2M_T(0) = 2M$ once all matter has fallen into the black hole.

Notice that Figures 7 and 8 have two very unusual properties. First, the event horizon increases in areal radius and then decreases as the black hole emits monopole radiation. This violates the famous area theorem, or second law of black hole dynamics, due to Hawking[31], which does not permit the surface area of an event horizon to decrease over time. Secondly, during the time in which the area of the event horizon decreases, the event horizon lies *inside* the apparent horizon. This is more easily seen in Figure 9, in which the apparent horizons and event horizons for three of the cases from Figures 7 and 8 are shown on the same plot. In other words, it possible for an observer to penetrate the apparent horizon, determine that he is in a region of trapped surfaces by shining a flashlight, and then escape to infinity†. This violates the well-known apparent horizon theorem of Hawking and Ellis[31], which requires an apparent horizon to lie inside or coincide with an event horizon.

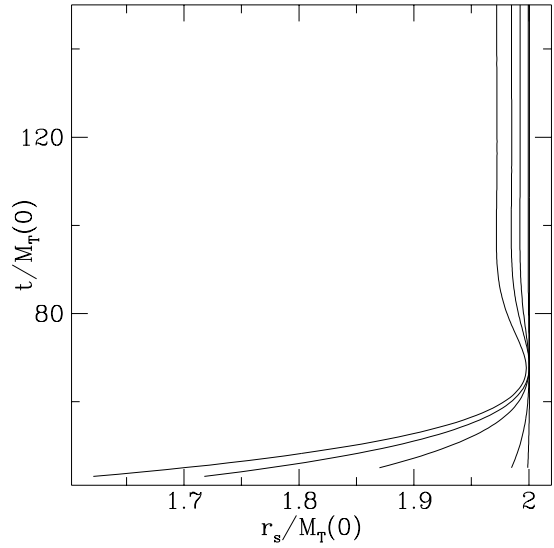

**Figure 7**  Spacetime diagram showing location of apparent horizons of black holes resulting from Oppenheimer-Snyder collapse with five different values of $\omega$. Horizons are plotted for, from left to right, $3 + 2\omega = 3, 5, 10, 100$, and $1000$.

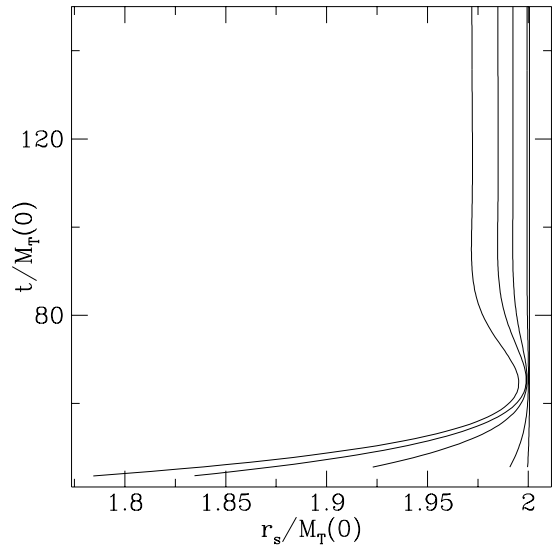

**Figure 8**  Spacetime diagram showing location of event horizons of black holes resulting from Oppenheimer-Snyder collapse with five different values of $\omega$. Horizons are plotted for, from left to right, $3 + 2\omega = 3, 5, 10, 100$, and $1000$. Since $r_s$ is the areal radius, the area of the event horizon is seen to *decrease* at times during the collapse.

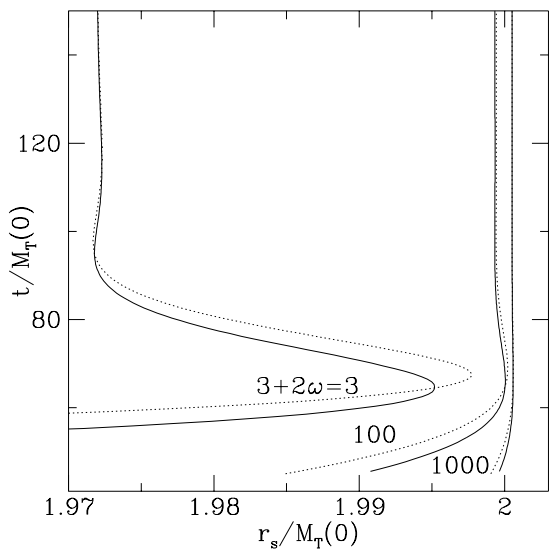

**Figure 9**  Spacetime diagram showing location of apparent horizons (dotted lines) and event horizons (solid lines) of black holes resulting from Oppenheimer-Snyder collapse with three different values of $\omega$, during the epoch in which the black holes emit monopole radiation. For $3 + 2\omega = 1000$, the horizons are located at $r_s$ slightly larger than $2M$ at late times because of numerical errors in the initial data.

According to our numerical results, *the area theorem and the apparent horizon theorem are violated in Oppenheimer-Snyder collapse for all values of $\omega$*. This violation occurs in vacuum, during the dynamical epoch in which the black hole is radiating mass. To see why this is possible, recall that most global black hole theorems, including the area and apparent horizon theorems, assume the null energy condition (1.2). Using the metric (4.1), we can evaluate the quantity $R_{ab}l^a l^b$ for the radially outgoing null vector

$$l^a = \left(\frac{1}{\alpha}, \frac{1}{A} - \frac{\beta}{\alpha}\right), \tag{5.3}$$

which is obtained by summing the unit normal vector $n^a$ and the outgoing radial spatial unit vector. Using the field equations (2.5) and (2.6), the metric (4.1), and the maximal slicing condition (4.2), we can write

$$R_{ab}l^a l^b = \frac{8\pi}{\phi}\left(\rho + S^r{}_r + \frac{2S_r}{A} - \frac{T}{3+2\omega}\right) + \frac{\omega}{\phi^2}\left(\Pi - \frac{\Phi}{A}\right)^2$$
$$- \frac{\Pi K^r{}_r}{\phi} - \frac{2\Pi_{,r}}{A\phi} + \frac{2\Phi K^r{}_r}{A\phi} + \frac{2}{A^2\phi}\left(\Phi_{,r} + \frac{\Phi}{r}\right), \tag{5.4}$$

---

† Thus, it is possible for an observer near a collapsing object to determine whether the collapse is governed by BD or GR. Only in the former case can he live to tell friends outside the black hole of his discovery.

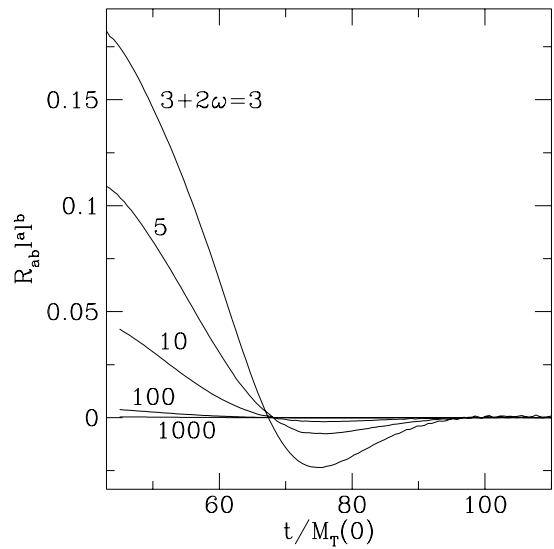

**Figure 10** The quantity $R_{ab}l^a l^b$ in units of $[M_T(0)]^{-2}$ at the apparent horizon versus time, for the black holes shown in Figure 7. The null energy condition is violated where this quantity is negative.

where $\rho$, $S^r_r$, $S_r$, and $T$ are matter variables defined in Section II.B of Paper I, $\Pi$ and $\Phi$ are derivatives of the scalar field $\phi$ defined by Eqs. (I.2.27) and (I.2.28), and $K^r_r$ is a component of the extrinsic curvature. The first term in Eq. (5.4) containing matter variables vanishes identically in vacuum. The second term, the only one involving $\omega$, results from the stress-energy of the scalar field. This term, which is what one would get from a Klein-Gordon field in general relativity, is manifestly nonnegative for $\omega \geq 0$. The final four terms result from the last two terms in the field equation (2.6). These terms are not necessarily positive, and in fact they cause the null energy condition to be violated during Oppenheimer-Snyder collapse. This can be seen in Figure 10, in which we plot $R_{ab}l^a l^b$ versus time for five values of $\omega$. Notice that the time during which $R_{ab}l^a l^b < 0$ corresponds to the time during which both the area theorem and apparent horizon theorem are violated in Figure 9.

We note that even in *linearized* Brans-Dicke theory, the null energy condition is violated in vacuum whenever the value of the scalar field dips below unity, independent of $\omega$. We show this by inserting the weak-field relations (I.2.45), (I.2.37), (I.2.38), (I.2.46) and (I.2.49) into Eq. (5.4) and working to first order in small quantites. We obtain

$$R_{ab}l^a l^b = \frac{2f(t-r)}{r^3}, \qquad (5.5)$$

where $f$ is the function appearing in Eqs. (5.1).

The fact that Brans-Dicke spacetimes can violate the area theorem and the apparent horizon theorem may come as a surprise to some readers because in the Einstein representation, Brans-Dicke theory in vacuum is simply general relativity with a Klein-Gordon scalar field. In this representation, the null energy condition (3.1) is always satisfied in vacuum for $\omega \geq -3/2$, as one can verify from Eq. (2.9). Thus, if one works in the Einstein representation, one concludes that the area and apparent horizon theorems must hold.

To see why this implies no contradiction, recall that one transforms to the Einstein representation by multiplying the metric by $G_0\phi$ and the stress-energy tensor by $G_0\phi^{-1}$ at each point in spacetime while leaving the coordinates invariant. Here $G_0$ is the arbitrary constant appearing in Eqs. (2.7), which we set to unity. Because this conformal transformation changes neither the coordinates of a given spacetime event nor the coordinate paths of light rays, it also does not affect whether a particular light ray at a given location in spacetime will escape to infinity or be pulled into the black hole. Therefore, the spacetime location of the event horizon is the same whether calculated in the Einstein representation or in the Brans-Dicke representation. However, the *surface area* of this event horizon will be different in the two representations, because the area of an $r = $ constant surface in the Einstein representation is $\phi$ times the area of this same $r = $ constant surface in the BD representation.

Unlike the event horizon, the apparent horizon depends not only on the trajectories of light rays, but also on how the area of a bundle of light rays changes in time. Because of this dependence on area, there exists two distinct surfaces that may be called apparent horizons: the physical apparent horizon that one calculates using the Brans-Dicke representation and the unphysical apparent horizon that one calculates using the Einstein representation. The equation describing the location of the physical apparent horizon, derived in Section II.F of Paper I, is

$$\frac{1}{r} + \frac{A_{,r}}{A} = -\frac{1}{2} A K^r_{\ r}, \tag{5.6}$$

where we have assumed maximal time slicing and the spatially isotropic gauge. In contrast, the unphysical apparent horizon obeys

$$\frac{1}{r} + \frac{A_{,r}}{A} = \frac{1}{2}\left(-AK^r_{\ r} + \frac{\Pi A}{\phi} - \frac{\Phi}{\phi}\right). \tag{5.7}$$

This equation can be obtained by the same method used in Paper I, with the substitution $\mathcal{A} \to \phi\mathcal{A}$ in Eq. (I.2.71).

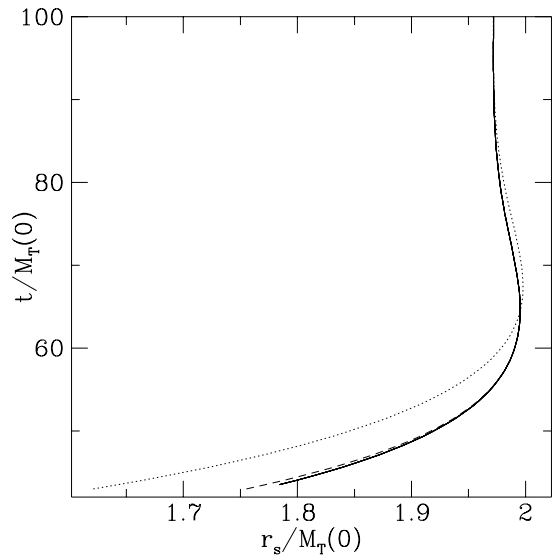

**Figure 11**   Spacetime diagram showing location of the apparent horizon (dotted line), event horizon (solid line), and unphysical Einstein representation apparent horizon (dashed line) of the black hole resulting from Oppenheimer-Snyder collapse with $\omega = 0$.

Both of these apparent horizons are shown in Figure 11 for the case $\omega = 0$. The unphysical apparent horizon always lies inside or coincides with the event horizon, in agreement with the apparent horizon theorem, while the physical apparent horizon crosses the event horizon. This is because the null energy condition (3.1) in the Einstein representation is satisfied, but the same condition (1.2) in the Brans-Dicke representation is not.

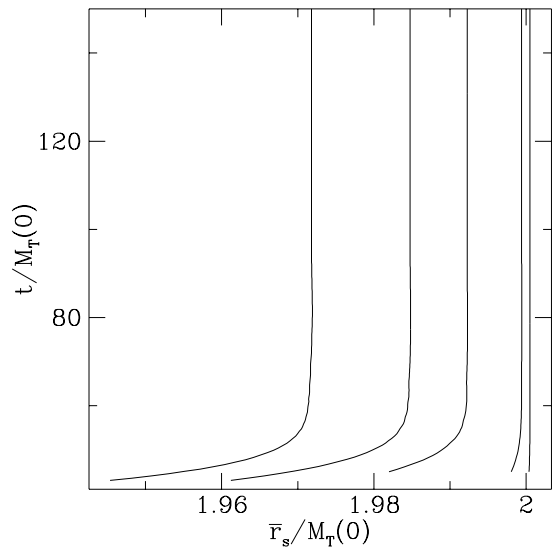

**Figure 12**  Spacetime diagram showing location of the unphysical Einstein representation apparent horizon for five values of $\omega$. On the horizontal axis is the areal radius $\bar{r}_s = \phi^{1/2} r_s$ as measured in the Einstein representation. Horizons are plotted for, from left to right, $3 + 2\omega = 3$, 5, 10, 100, and 1000. For $3 + 2\omega = 1000$, the horizon is located at $\bar{r}_s$ slightly larger than $2M$ at late times because of numerical errors in the initial data.

In Figure 12, we show the unphysical apparent horizons of black holes with five different values of $\omega$, plotted in terms of the areal radius $\bar{r}_s = \phi^{1/2} r_s$ as measured in the Einstein representation. The area of the unphysical apparent horizon never decreases, except for small numerical errors. Because the event horizon coincides with the unphysical apparent horizon after $t = 50 M_T(0)$ (Figure 11), we conclude that as measured in the Einstein representation, the area of the event horizon never decreases, in agreement with Hawking's theorem.

We conclude that in the Einstein representation of the theory, black holes in BD are the same as those in GR. However, in the physical Brans-Dicke representation, *dynamical* black holes can behave quite differently than their general relativistic counterparts.

### B. Oppenheimer-Snyder Collapse for $\omega < -2$

For completeness, we also examine Oppenheimer-Snyder collapse for negative values of $\omega$. We begin with an initially stationary uniform particle distribution with tensor mass $M_T(0)$ and areal radius $R_s = 10 M_T(0)$. We only consider $\omega < -2$ because of numerical subtleties in handling spacetimes with $-2 \lesssim \omega \lesssim 0$. These subtleties stem from the singular behavior of the Brans-Dicke equations at $\omega = -3/2$.

As is the case with positive values of $\omega$, we find that for negative $\omega$ the collapse results in a black hole rather than a naked singularity. In addition, the black hole approaches the Schwarzschild solution with constant scalar field at late times.

Since the scalar field couples negatively to matter, the scalar mass, tensor mass, and active gravitational mass now *increase* as the black hole radiates. Gravitational waveforms are similar to those for the $\omega > 0$ cases, except that the overall sign of the quantity $\phi - 1$ has changed. Figure 13 shows $\mathcal{M}_S(80 M_T(0))$ as a function of time for five different negative values of $\omega$. The scalar mass is initially negative, and then increases to zero as the black hole reaches final equilibrium.

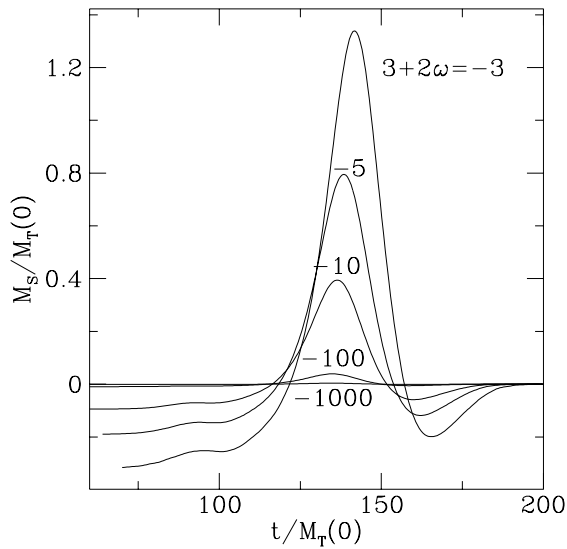

**Figure 13** Scalar mass function $\mathcal{M}_\mathrm{S}(r)$ versus time for Oppenheimer-Snyder collapse, calculated at $r = 80 M_\mathrm{T}(0)$ from Eq. (I.2.64.b) and shown for five values of $\omega$. The quantity $\mathcal{M}_\mathrm{S}(r)$ is well-defined and unique only for stationary systems, so only the initial (negative) and final values of $\mathcal{M}_\mathrm{S}(r)$ are physically meaningful, and are equal to the scalar mass $M_\mathrm{S}$. The scalar mass is zero in the final state.

The apparent horizon, event horizon, and unphysical Einstein representation apparent horizon for several collapse scenarios with negative $\omega$ are shown in Figure 14. Notice that the apparent horizon is only *inside* the event horizon and the event horizon only *increases* in area for a short time interval during the epoch in which the black hole emits radiation. Furthermore, the unphysical Einstein representation apparent horizon is always *outside* or coincident with the event horizon. This is opposite to the case for positive $\omega$, because the Einstein representation null energy condition (3.1) is always violated in vacuum for $\omega < -3/2$, and the Brans-Dicke representation null energy condition (1.2) is only satisfied in vacuum for a short time while the black hole radiates mass.

## VI. CONCLUSION

Using a new numerical code, we have demonstrated that Oppenheimer-Snyder collapse in Brans-Dicke theory results in black holes rather than naked singularities, at least for $|3 + 2\omega| \geq 3$. We have shown that dynamical black holes in Brans-Dicke theory can behave quite differently than those in general relativity: because the null energy condition (1.2) is violated even in vacuum spacetimes with positive values of $\omega$, the apparent horizon of a black hole can pass *outside* the event horizon, and the surface area of the event horizon can *decrease* over time. For nonnegative $\omega$, this behavior occurs while the black hole, soon after the initial collapse, is radiating its scalar mass to infinity. If $\omega < 0$, the opposite behavior occurs: only during a small time interval while the black hole radiates is the event horizon located outside the apparent horizon and its surface area increasing. Once the black hole reaches final equilibrium, the Brans-Dicke scalar field is constant, and the spacetime metric is the Schwarzschild solution. Thus, in the final stationary state, a spherical black hole in Brans-Dicke theory is indistinguishable from those of general relativity.

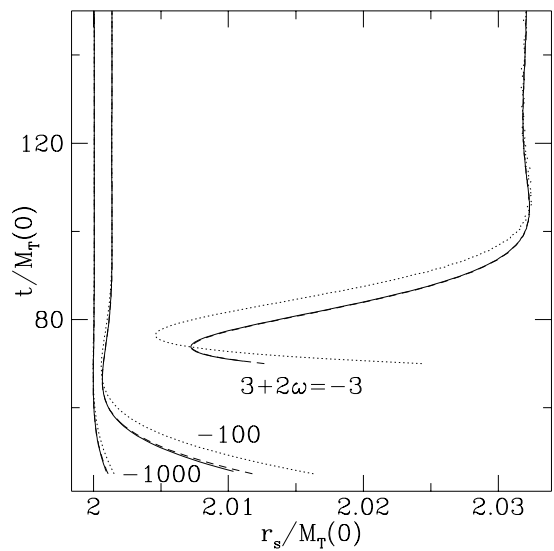

**Figure 14** Spacetime diagram showing location of apparent horizons (dotted lines) event horizons (solid lines), and unphysical Einstein representation apparent horizons (dashed lines) of black holes resulting from Oppenheimer-Snyder collapse with three different values of $\omega < 0$, during the epoch in which the black holes emit monopole radiation.

## ACKNOWLEDGMENTS

We would like to thank T. Baumgarte for helpful discussions. This work has been supported in part by National Science Foundation grants AST 91-19475 and PHY 90-07834 and the Grand Challenge grant NSF PHY 93-18152/ ASC 93-18152 (ARPA supplemented).